%% file: 9301293.tex
\documentstyle[11pt]{article}

\newif\ifIncludeFigs\IncludeFigsfalse
\newif\ifTwoUp\TwoUpfalse
\newif\ifEPSF\EPSFfalse

\IncludeFigstrue
\input epsf

\ifIncludeFigs
\ifx\epsfbox\undefined
  \EPSFfalse
\else
  \EPSFtrue
\fi
\fi
\ifx\endfloat\notincluded\else
  \nomarkersintext
\fi

\catcode`@=11
\def\caption{\@dblarg\aux@caption}

\def\aux@caption[#1]#2{
   \parindent 20pt \par
      {\refstepcounter\@captype \@caption{\@captype}[#1]{#2}}}

\long\def\@caption#1[#2]#3{\par\addcontentsline{\csname
  ext@#1\endcsname}{#1}{\protect\numberline{\csname
  the#1\endcsname}{\ignorespaces #2}}\begingroup
    \@parboxrestore
    \small\sl
    \@makecaption{\csname fnum@#1\endcsname}{\ignorespaces #3}\par
  \endgroup}

\def\maketitle{
 \begingroup
 \def\thefootnote{\fnsymbol{footnote}}
 \def\@makefnmark{\hbox
 to 0pt{$^{\@thefnmark}$\hss}}
 \if@twocolumn
 \twocolumn[\@maketitle]
 \else 
 \global\@topnum\z@ \@maketitle \fi
 \thispagestyle{empty}
 \setcounter{page}{0}
 \@thanks
 \endgroup
 \setcounter{footnote}{0}
 \let\maketitle\relax
 \let\@maketitle\relax
 \gdef\@thanks{}\gdef\@author{}\gdef\@title{}\let\thanks\relax}

\def\paperid#1{\gdef\@paperid{#1}}

\def\@maketitle{

 \@makepub
 \vskip 2em \begin{center}
 { \Large \bf \@title \par}
 \vskip 1.5em {\large \lineskip   .5em
 \@authoraddress
 }
 \end{center}
 \par
 \vskip 1.5em
}

\def\@makepub{{
  \centering
  \makebox[\textwidth]{
    \parbox[t]{0.25\textwidth}{\begin{flushleft}%
      {\small\@pubdate}\end{flushleft}}
    \hfil
    \parbox[t]{0.5\textwidth}{\begin{center}%
      {\small \@publabel}\end{center}}
    \hfil
   \parbox[t]{0.25\textwidth}{\begin{flushright}{\small
    \@pubnumber}\end{flushright}}
  }
}}

\gdef\@publabel{\hfil}
\gdef\@pubdate{Jan 1, 1999}
\gdef\@pubnumber{EFI-??-??}

\long\def\pubdate#1{\gdef\@pubdate{#1}}
\long\def\pubnumber#1{\gdef\@pubnumber{#1}}
\long\def\publabel#1{\gdef\@publabel{#1}}

\def\abstract{\if@twocolumn
\section*{ABSTRACT}
\else \small
\begin{center}
{
ABSTRACT\vspace{-.5em}\vspace{0pt}}
\end{center}
\quotation
\fi%
}

\def\endabstract{\if@twocolumn\else\endquotation\fi}

\def\pacs#1{\par %
  \bgroup
  \hsize\columnwidth \parindent0pt
  \if@twocolumn\else\leftskip=0.10753\textwidth \rightskip\leftskip\fi
 \ifdim\prevdepth=-1000pt \prevdepth0pt\fi
 \dimen0=-\prevdepth \advance\dimen0 by20pt\nointerlineskip
  \vbox to28pt{\small\vrule height\dimen0 width0pt\relax%
	PACS: #1\vfill}
  \egroup
  \if@twocolumn\vskip1pc\fi
  \newpage
}

\gdef\@author{Nobody}
\gdef\@authoraddress{}

\def\@makeauthor{
  {\def\and{\smallskip {\normalsize \rm and\smallskip}}
  {\zerospfalse \centering \large \@author}
  }
}

\def\author#1{\expandafter\def\expandafter\@authoraddress\expandafter
  {\@authoraddress %
  {
  \dimen0=-\prevdepth \advance\dimen0 by23pt
  \nointerlineskip
  \rm\centering
  \vrule height\dimen0 width0pt\relax\ignorespaces#1%
      \\[\baselineskip] 
  }%
  }%
}

\def\address#1{\expandafter\def\expandafter\@authoraddress\expandafter
  {\@authoraddress{\small\it\centering \baselineskip 1.3\baselineskip
\ignorespaces#1 \par
  }}
}

\def\thebibliography#1{\newpage
\section*{References\markboth{REFERENCES}{REFERENCES}}
\addcontentsline{toc}{section}{References}\labelsep1.0em\list
  {\arabic{enumi}.}{\settowidth\labelwidth{#1.}%
  \leftmargin\labelwidth
    \advance\leftmargin\labelsep\usecounter{enumi}}}

\let\acknowledgements=\acknowledgement
\def\noteadded{\msubsection*{Note Added}}
\def\msubsection{\@startsection{subsection}{2}{0.25em}%
  {4.5ex plus 1ex minus .2ex}{1.0ex plus .2ex}{\normalsize\bf}}
\def\twoup{
        \twocolumn\sloppy\flushbottom\parindent 2em
        \leftmargini 2em\leftmarginv .5em\leftmarginvi .5em
        \oddsidemargin -.5in    \evensidemargin 0in
        \columnsep .4in \footheight 0pt
        \textwidth 10in \topmargin  -.4in
        \headheight 0pt \topskip 0in
        \textheight 6.9in \footskip 30pt
        \def\@oddfoot{\hfil\thepage\hfil\addtocounter{page}{1}
                \hspace{\columnsep}\hfil\thepage\hfil}
        \let\@evenfoot\@oddfoot \def\@oddhead{} \def\@evenhead{}
}

\newdimen\slashraise \slashraise=1pt
\mathchardef\fslash="0236

\def\slash@char#1#2{\setbox0=\hbox{$#2$}           
   \dimen0=\wd0                                 
   \dimen2=-\dp0 \advance\dimen2 by \slashraise
   \setbox1=\hbox{$\m@th#1\mkern-13mu\fslash$}
	 \dimen1=\wd1               
   \ifdim\dimen0>\dimen1                        
      \rlap{\hbox to \dimen0{\hfil\raise\dimen2\box1\hfil}}%
      #2                                        
   \else                                        
      \rlap{\hbox to \dimen1{\hfil$#2$\hfil}}   
      \raise\dimen2\box1                              
   \fi}                                         %

\def\slashchar#1{\mathrel{\mathpalette\slash@char#1}}
\let\slsh=\slashchar

\catcode`@=11

\ifx\@ove@rcfont\undefined
  \ifx\citen\undefined
     \def\citen#1{\begingroup \def\@cite##1##2{{##1}}%
	\@citex[]{#1}\endgroup}
  \fi
\else
  \def\@cite#1{$\@ove@rcfont\m@th^{[\hbox{#1}]}$}
\fi

\def\wider#1{\dimen1=#1 \divide \dimen1 by 2
\advance \textwidth by #1 \advance \oddsidemargin by -\dimen1
  \advance \marginparwidth by -\dimen1  \evensidemargin\oddsidemargin
  \hsize\textwidth}
\def\partder#1#2{{\partial #1\over\partial #2}}

\def\efiphys{\address{%
  Enrico Fermi Institute and Department of Physics\\
              University of Chicago, Chicago, Illinois 60637}}

\def\bar{\overline}
\def\eqb{\begin{equation}}
\def\eqe{\end{equation}}

\def\hardfill#1{\vrule depth \z@ height\z@ width #1}
\def\mpty{\mbox{}}

\def\pslsh{\slsh{p}}

\def\kslsh{\slsh{k}}

\def\Order{O}

\def\Im{{\rm Im}}
\def\Mu#1{m_{u_{#1}}}
\def\Md#1{m_{d_{#1}}}

\def\Amp{{\cal A}}

\def\GammaIR{\Gamma^{\rm IR}}
\def\Sigmab{\bar\Sigma}
\def\ecm{e\,{\rm cm}}

\ifTwoUp
  \twoup
\fi

\def\Month{\ifcase\month\or
 January\or February\or March\or April\or May\or June\or
 July\or August\or September\or October\or November\or December\fi
}
\newcount\hr
\newcount\min
\newcount\xx
\def\TimeStamp{\hr=\time \divide \hr by 60
        \xx=\hr \multiply\xx by 60
	\min=\time \advance \min by -\xx
	\hbox{\ifnum \hr < 10 0\fi\number\hr :\number\min}}

\pubnumber{EFI-93-01\\ hep-ph/9301293}
\pubdate{January, 1993}
\begin{document}
\begin{titlepage}
\title{The Electric Dipole Moment of the W and Electron in the
	Standard Model}
\author{Michael J.~Booth%
 \footnote{Electronic address: booth@yukawa.uchicago.edu}
}
\efiphys
\maketitle
\begin{abstract}
I show that the electric
dipole moment of the W-boson $d_W$ vanishes to two loop order in the
standard Kobayashi-Maskawa Model of $CP$ violation.  The argument is a
simple generalization of that used to show the vanishing of the quark
electric dipole moment.  As a consequence, the electron electric
dipole moment vanishes to {\em three} loop order.
Including QCD corrections may give a
non-vanishing result; I estimate $d_W \approx 8\cdot 10^{-31}\ecm$,
which induces an electron EDM $d_e \approx 8 \cdot 10^{-41}\ecm$,
considerably smaller than a previous calculation.
\end{abstract}
\end{titlepage}
\section{Introduction}

Any model of $CP$ violation will in general induce ---
through loop effects --- $P$ and $T$ violating electric
dipole moments (EDMs) for elementary particles, including
the $W$-boson.
The $P$ and $T$ violating interaction of the $W$ boson with
a photon can be described by the effective Lagrangian\cite{Peccei}
\eqb
\label{lagrangian}
{\cal L}^{CP}_{W\gamma} = i\lambda_1 \tilde F_{\mu\nu}W^\mu W^\nu +
	i{\lambda_2\over M_W^2} \tilde F_{\mu\nu}W^\mu{}_\sigma
	W^{\sigma\nu}.
\eqe
Here $\tilde F_{\mu\nu} = \frac{1}{2}\epsilon_{\mu\nu\alpha\beta}
	F^{\alpha\beta}$,
	$W_{\mu\nu} = \partial_\mu W_\nu - \partial_\nu W_\mu$
and $W_\nu$ is the $W$-boson field%
\footnote{
  As an aside, note that the first term in
  eqn. (\ref{lagrangian}) is not $SU(2)$ gauge invariant, so it
  should be accompanied by operators containing the Higgs field and its
  coefficient should be proportional to $SU(2)$ breaking terms ---
  it would not be present in an unbroken theory.
}.
In terms of the Lagrangian eqn (\ref{lagrangian}),
the coefficient of the W-boson EDM (WEDM) is then given by $d_W =
	(\lambda_1 + \lambda_2/M_W^2)(e/ 2 M_W)$.

Although it is possible in principle to measure the WEDM directly, for
example in scattering experiments\cite{Queijeiro}, it will also
contribute
through loops in lower energy phenomena.
It was proposed years ago as an explanation of $CP$ violation
in $K^0\ \bar{K}^0$ mixing\cite{Salzman^2}.  In most models, the WEDM
(and related operators) give the dominant contribution
to the the electron EDM.  This is believed to be the case in
the Standard Kobayashi-Maskawa Model (KM model) of $CP$ violation,
where the contribution of the WEDM to the electron EDM has been
calculated to be $d_e \simeq 10^{-38}\ecm$\cite{Hoogeveen},
although this is in conflict with an estimate of $d_W$%
\cite{ChangWEDM} which is itself on the order of $10^{-38}\ecm$.

In this note I will show that in fact $d_W$ vanishes to
two-loop order in the KM model.  This in turn implies that
$d_e$ vanishes to three-loop order, contradicting the result
of ref. (\citen{Hoogeveen}). The remainder of this paper
is organized as follows:  in section two I
review the mechanics of $CP$ violation in the KM model. In section
three I demonstrate the vanishing of $d_W$ and in the following
section I estimate the three-loop QCD contribution to
$d_W$ and $d_e$.
Finally, in section five I conclude.


\section{The KM Model}

Unitarity of the CKM matrix and rephasing invariance imply that
all $CP$ violating effects in the standard model are governed by
the quantity $\Phi_{u_1 d_1}^{u_2 d_2} =
 \Im(V_{u_1d_1} V^*_{u_2d_1} V_{u_2d_2} V^*_{u_1d_2})$.
Here $V_{ud}$ is KM matrix element and $u_i, d_j$ are arbitrary
up and down-type quarks.
$\Phi$ has two important properties\cite{Jarlskog,Isi}: it is
antisymmetric in $(u_1, u_2)$ and $(d_1, d_2)$ and the sum over any
index, {\it eg}. $u_1$, vanishes.  For three families there is the
particularly simple relation\cite{Jarlskog}
$\Phi_{u_1 d_1}^{u_2 d_2} = J\, \sum_{\gamma k} \epsilon_{u_1 u_2
\gamma} \epsilon_{d_1 d_2 k}$.
Any $CP$ violating amplitudeis given by summing $\Phi$ together with
the Feynman diagrams
$\Amp_{u_1 u_2}^{d_1 d_2}$
for each configuration of quarks that contribute to the
process in question.  $\Amp$ is a function of quark masses and
possibly KM angles (not phases).  In the case where the weak
interactions
occur along a single quark line --- as is true for the diagrams
which generate $d_W$ --- there is no additional dependence on the
KM angles so that one has
$\Amp_{u_1 u_2}^{d_1 d_2} = \Amp(\Mu1^2,\Mu2^2,\Md1^2,\Md2^2)$.
Since the weak interactions in the KM model are purely left handed,
quark masses enter the diagrams quadratically, through the
denominators of the rationalized propagators%
\footnote{
  Strictly speaking, this is true only in the unitary gauge.  For
  other gauge choices, mass dependence will also enter through the
  unphysical Higgs vertices.  However, it is possible to arrange
  the calculation so that these masses drop out.
}.
Because of the antisymmetry of $\Phi$, only those
parts of Feynman diagrams which are not symmetric under the exchange
of up or down quark masses will contribute to $CP$ violation. This
anti-symmetrization, which is the GIM mechanism,
leads to cancellations between the contributions of
different quarks and is responsible for suppressing most $CP$
violating effects in the KM model\cite{BBS}.

\section{The WEDM in the KM Model}
The two loop contributions to the WEDM in the KM model are
shown in figure~1.  An additional set of five is
generated by interchanging the roles of the up and down quarks in the
loop, for a total of ten diagrams.  I will only consider
the first set, since the treatment of the second set is exactly
parallel.

\ifIncludeFigs
\begin{figure}
     \ifEPSF
	\centerline{\epsfbox{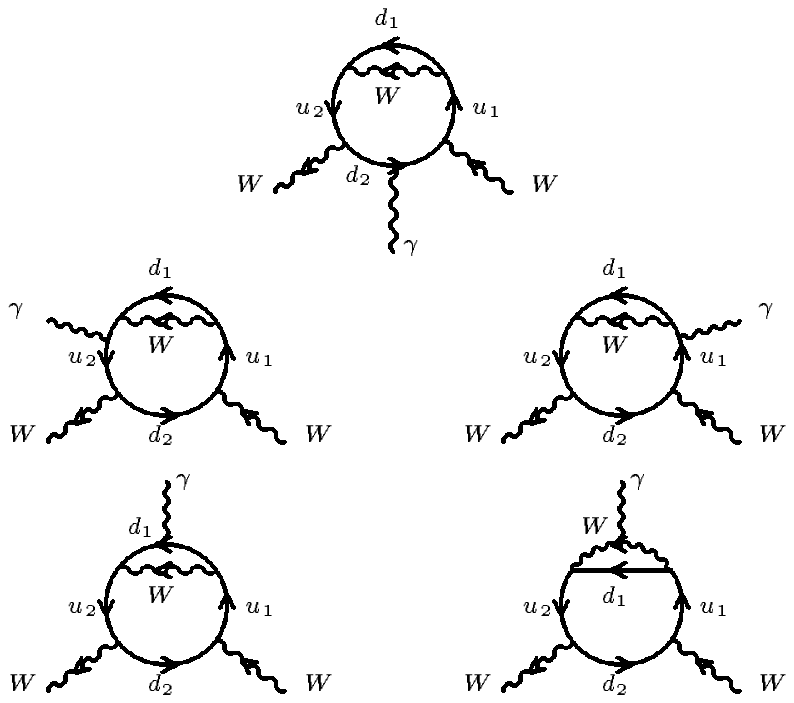}}%
     \else
       \begin{center}
         \input wedm-fig1
       \end{center}
     \fi
   \caption{\label{wedm}
   The two-loop contributions to $d_W$.
   }
\end{figure}
\fi

It is easy to eliminate from further consideration
the diagram where the photon attaches to
the bottom of the loop in figure \ref{wedm}:
the photon momentum $k$ appears only in the down quark line, so
the two up-quark propagators have the same momentum dependence.
Consequently, the
diagram is symmetric in the up-quark masses and as discussed
earlier will not contribute to the WEDM; it will be clear from the
renormalized form of the self-energy that this statement remains true
after renormalization.

Instead of studying the four remaining diagrams directly, consider a
simpler set of diagrams obtained by cutting the $d_2$ quark line
in figure \ref{wedm}.
They are shown in figure \ref{1loop}.

\ifIncludeFigs
\begin{figure}
   \ifEPSF
     \centerline{\epsfbox{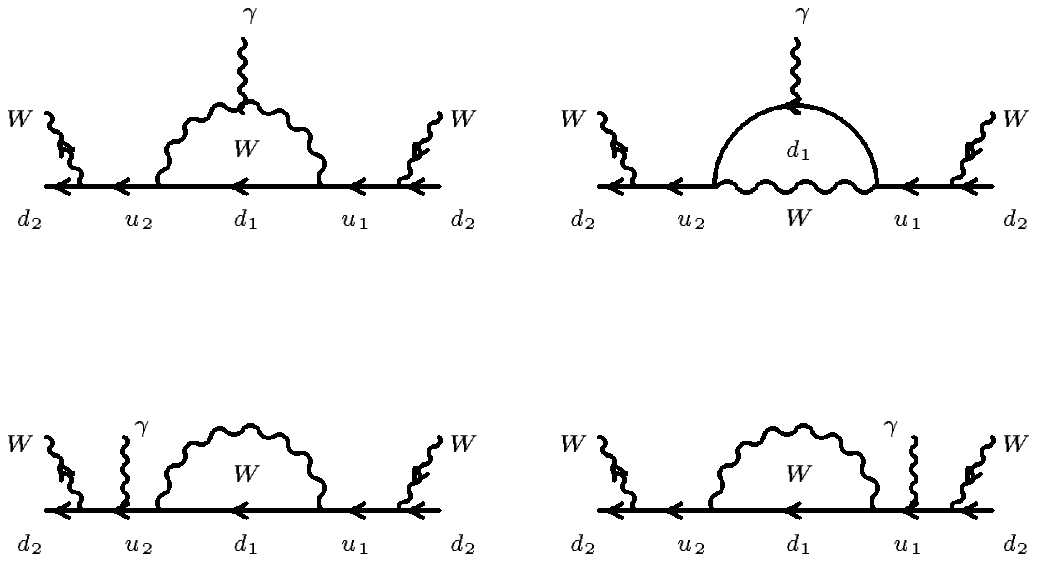}}
   \else
     \begin{center}
       \input wedm-fig2
     \end{center}
   \fi
    \caption{\label{1loop} Vertex sub-graphs of the diagrams of
	figure 1. }
\end{figure}
\fi

The diagrams of figure \ref{1loop}
combine to give (keeping only
the chiral projectors from the W-vertices)
\begin{equation}
\Amp_\mu = L {1\over \pslsh - {\kslsh\over2} - \Mu2}
  \Gamma_\mu(p, k)
  {1\over \pslsh + {\kslsh\over2} - \Mu1} R
\end{equation}
where $\Gamma_\mu$ is the complete vertex function defined by
\begin{eqnarray}
\Gamma_\mu(p,k) &=&
  \gamma_\mu
    {1\over\pslsh-{\slsh k\over2} -\Mu2}\Sigma(p-k/2) +
  \mpty \Sigma(p+k/2){1\over\pslsh+{\slsh k\over2} -\Mu1}\gamma_\mu
  \nonumber \\
&&\mpty +\GammaIR_\mu(p,k),
\end{eqnarray}
$\Sigma$ is the renormalized self-energy function and $\GammaIR$ is the
irreducible vertex function.  These have been computed many times
before%
\cite{Shabalin,ShabalinV,Others} and their detailed forms
are not required.  However,
because the results are not generally well known, I will sketch
the calculation.
Since the EDM is a static effect, I will
expand the vertex in powers of the photon momentum $k$, keeping only
the first term\footnote{
  An equivalent analysis of the vertex function using a
  different approach has recently been presented in
  ref. \citen{Me}.
}.

After renormalization, the self-energy $\Sigma$ takes the form
\begin{equation}
\Sigma_{u_2 u_1} =
        (\pslsh - m_{u_2})\bar\Sigma_{u_2u_1}(\pslsh - m_{u_1})
\end{equation}
with
\begin{equation}
\bar\Sigma = F_0^{(1)}(p^2) ( \pslsh R + m_{u_2}R + m_{u_1}L )
 + F_0^{(2)}(p^2)\pslsh L.
\end{equation}
Here $F_0^{(1)}$ and $F_0^{(2)}$ are 
\begin{eqnarray}
F_0^{(1)} & = & {p^2 F_0 + \Mu1\Mu2 X_0\over
	(p^2 - \Mu1^2) (p^2 -\Mu2^2)},\\
F_0^{(2)} & = & {\Mu1\Mu2F_0 + (\Mu1^2 + \Mu2^2 - p^2)X_0\over
	(p^2 - \Mu1^2) (p^2 -\Mu2^2)}
\end{eqnarray}
and
\begin{eqnarray}
F_0 &= &f(p^2) -
	{\Mu1^2f(\Mu1^2) - \Mu2^2 f(\Mu2^2) \over \Mu1^2 - \Mu2^2},\\
X_0 & = & \Mu1\Mu2{f(\Mu1^2) - f(\Mu2^2) \over \Mu1^2 - \Mu2^2}.
\end{eqnarray}
Finally, $f(p^2)$ is the function which occurs in the unrenormalized
self energy, $\Sigma_0(p) = f(p^2)\pslsh L$.
Similarly, the irreducible vertex has the form
\begin{eqnarray}
\GammaIR_\mu(p,k) = -Q_u \partder{}{p_\mu} \Sigma +
 f_2 \{\pslsh, \sigma_{\mu\nu} k^\nu \}L,
\end{eqnarray}
where $f_2(p^2)$ does not depend on the down-quark masses.

In terms of the above functions, $\Amp_\mu$ has the relatively
simple form
(to \Order(k))
\begin{eqnarray}
\Amp_\mu(p,k) &=& \{\pslsh, \sigma_{\mu\nu} k^\nu \} R\,
{p^2 f_2(p^2) + p^2 F_0^{(1)} - \Mu1\Mu2 F_0^{(2)} \over
	(p^2 - \Mu1^2)(p^2 - \Mu2^2)} \nonumber\\ &&\mpty
- Q_u \partder{}{p_\mu} L\Sigmab(p)R.
\end{eqnarray}
Recalling the expression for $\Sigmab$ one sees that
$\Amp_\mu$ is a symmetric function of $\Mu1$ and $\Mu2$.
It is this symmetry which is at the heart of all the vanishing
two loop effects in the KM model.
It follows that the diagrams of figure 1 combine to produce
a symmetric function of the up-quark masses
so that
as I intended to show,
the W-boson EDM vanishes to two loops in the Standard Model.

\section{Estimates}
In order to obtain a non-zero WEDM, it is necessary to destroy the
symmetric way in which the quark propagators enter the diagrams.
This can be accomplished by including QCD loop corrections.
The resulting
3-loop diagrams are rather difficult to calculate. One way to
simplify these calculations would be to perform an effective
lagrangian expansion, including gluonic operators, and then
integrate out the gluon fields.
However,
in view of the small expected size of the WEDM in the KM model,
a cruder estimate will suffice.

Because the GIM mechanism effectively cuts off the
the momentum integrals, it is reasonable to estimate them
by their infrared limits.  Gluon loops will typically
introduce logarithmic mass dependence.  This is also true of
the functions $f(p^2)$ and $f_2(p^2)$.  Moreover, the entire
diagram has a superficial logarithmic divergence. Consequently
I expect the mass dependence of the diagrams contributing to
$d_W$ to be logarithmic.
Because of the insensitivity of the
logarithm --- even for the widely separated scales in the KM model
--- the GIM cancellation will be weak and
I can obtain a reasonable (over-) estimate of $d_W$ by setting these
logs to 1.  I then obtain
\eqb
d_W \approx J \big( {1\over 16 \pi^2}\big)^2 \big({g_W^2 \over8}\big)^2
	\big({\alpha_s\over 4\pi}\big)
	\big({e\over 2 M_W}\big) \simeq 8 \cdot 10^{-30} \ecm.
\eqe
In ref. (\citen{ChangWEDM}) Chang {\it et al.} estimated the vanishing
two-loop contribution to be
\eqb
d_W = J \big({g_W^2\over 8 \pi^2}\big)^2 \big({e\over 2 M_W}\big)
	{m_b^4 m_s^2 m_c^2\over M_W^8}.
\eqe
However, this estimate is unduly pessimistic because it assumes the
quark mass dependence of the diagrams is polynomial, leading to
the small mass ratios in their estimate.

In order to estimate the electron EDM $d_e$, I use the
analysis of Marciano and Queijeiro\cite{Marciano}, who have updated the
original study by Salzman and Salzman\cite{Salzman^2}.
They calculate the contribution of
$d_W$ to $d_e$ and obtain the relation
\eqb
d_e \approx \big({g_W^2 \over 32 \pi^2} \big) \big({m_e\over M_W})
  \big[\ln{\Lambda^2\over M_W^2} +\Order(1)\big] \, d_W,
\eqe
where $\Lambda$ is a cutoff describing the scale of ``new physics''.
Setting the term in brackets to 1, I find for the KM model
\eqb
d_e \approx 8 \cdot 10^{-41}\ecm,
\eqe
which is considerably smaller than the value obtained by Hoogeveen
\cite{Hoogeveen}, whose calculation of $d_e$ contains the
graphs of figure \ref{wedm} as sub-graphs and should thus vanish.
\section{Conclusions}

The vanishing of the WEDM to two-loop order in the KM model
shows that the W-boson
is on the same footing as all other fundamental particles in the
KM model: none of them possess electric dipole moments to two-loop
order.  This was shown years ago for the quarks\cite{Shabalin,Donoghue}
and leptons\cite{Donoghue} and more recently it was shown that
the chromo-electric dipole moment of the gluon (the Weinberg operators)
also vanishes\cite{Me}.

In order to generate a non-vanishing EDM for these particles,
it is necessary to destroy the symmetric way in which the quark
propagators enter the diagrams.  This can be done by considering
higher-order operators, or at the cost of another loop
by including QCD corrections.  Either way, this leads
to an extra suppression of $CP$ violating effects in the standard
model.

\noteadded
After completing this work, I became aware of the work of
Khriplovich and Pospelov (ref. \citen{Khriplovich}) who also consider
this problem.  Inspired by the same work
(E.~P.~Shabalin, ref. \citen{Shabalin}) we reach the same conclusion
about the vanishing $d_W$ and $d_e$.  My work can be viewed as
an extension of theirs in so far as I obtain actual estimates
of the QCD loop contribution to $d_W$ and $d_e$.

\acknowledgements
I am grateful to Scott Willenbrock for bringing reference
\citen{Khriplovich} to my attention.
This work was supported in part by DOE grant
AC02 80ER 10587.

\def\jvp#1#2#3#4{#1~{\bf #2}, #3 (#4)}
\def\PR#1#2#3{\jvp{Phys.~Rev.}{#1}{#2}{#3}}
\def\PRD#1#2#3{\jvp{Phys.~Rev.~D}{#1}{#2}{#3}}
\def\PRL#1#2#3{\jvp{Phys.~Rev.~Lett.}{#1}{#2}{#3}}
\def\PLB#1#2#3{\jvp{Phys. Lett.~B}{#1}{#2}{#3}}
\def\NPB#1#2#3{\jvp{Nucl.~Phys.~B}{#1}{#2}{#3}}
\def\SJNP#1#2#3{\jvp{Sov.~J.~Nucl.~Phys.}{#1}{#2}{#3}}
\def\AP#1#2#3{\jvp{Ann.~Phys.}{#1}{#2}{#3}}
\def\PL#1#2#3{\jvp{Phys.~Lett.}{#1}{#2}{#3}}
\def\NuovoC#1#2#3{\jvp{Nuovo.~Cim.}{#1}{#2}{#3}}

\ifIncludeFigs\else
 \newpage
  \begin{figure}
     \caption{\label{wedm}
     The two-loop contributions $d_W$.
     }
  \end{figure}
  \begin{figure}
      \caption{\label{1loop} Vertex sub-graphs of the diagrams of
	figure 1. }
  \end{figure}
\fi

\end{document}

%% file: orig/wedm-fig1.tex
\font \wedm=wedm
\unitlength=1mm
\begin{center}
\begin{picture}(45,25)
  \put(0,0){\wedm b}
  \put(-1,0){$\scriptstyle W$} 
  \put(29,0){$\scriptstyle W$} 
  \put(10, 0){$\scriptstyle d_2$}
  \put(13, 9){$\scriptstyle W$}
  \put(11, 16){$\scriptstyle d_1$}
  \put( 5, 8){$\scriptstyle u_2$}
  \put(23, 8){$\scriptstyle u_1$}
  \put(13, -6){$\scriptstyle \gamma$} 
\end{picture}
\end{center}

\begin{picture}(45,25)
  \put(0,0){\wedm l}
  \put(-1,0){$\scriptstyle W$} 
  \put(29,0){$\scriptstyle W$} 
  \put(10, 0){$\scriptstyle d_2$}
  \put(13, 9){$\scriptstyle W$}
  \put(11, 16){$\scriptstyle d_1$}
  \put( 5, 8){$\scriptstyle u_2$}
  \put(23, 8){$\scriptstyle u_1$}
  \put(-1, 13){$\scriptstyle \gamma$} 
\end{picture}
\begin{picture}(45,25)
  \put(0,0){\wedm r}
  \put(-1,0){$\scriptstyle W$} 
  \put(29,0){$\scriptstyle W$} 
  \put(10, 0){$\scriptstyle d_2$}
  \put(13, 9){$\scriptstyle W$}
  \put(11, 16){$\scriptstyle d_1$}
  \put( 5, 8){$\scriptstyle u_2$}
  \put(23, 8){$\scriptstyle u_1$}
  \put(29, 13){$\scriptstyle \gamma$} 
\end{picture}

\begin{picture}(45,25)
  \put(0,0){\wedm m}
  \put(-1,0){$\scriptstyle W$} 
  \put(29,0){$\scriptstyle W$} 
  \put(10, 0){$\scriptstyle d_2$}
  \put(13, 9){$\scriptstyle W$}
  \put(11, 16){$\scriptstyle d_1$}
  \put( 5, 8){$\scriptstyle u_2$}
  \put(23, 8){$\scriptstyle u_1$}
  \put(16, 21){$\scriptstyle \gamma$} 
\end{picture}
\begin{picture}(45,25)
  \put(0,0){\wedm M}
  \put(-1,0){$\scriptstyle W$} 
  \put(29,0){$\scriptstyle W$} 
  \put(10, 0){$\scriptstyle d_2$}
  \put(13, 9){$\scriptstyle d_1$}
  \put(11, 16){$\scriptstyle W$}
  \put( 5, 8){$\scriptstyle u_2$}
  \put(23, 8){$\scriptstyle u_1$}
  \put(16, 21){$\scriptstyle \gamma$} 
\end{picture}

%% file: orig/wedm-fig2.tex
\font \wedm=wedm
\unitlength=1mm
\begin{picture}(50,20)
  \put(0,0){\wedm 2}
  \put(-3,6){$\scriptstyle W$}
  \put(41,6){$\scriptstyle W$}
  \put(19,3){$\scriptstyle W$} 
  \put(-3,-4){$\scriptstyle d_2$}
  \put(41,-4){$\scriptstyle d_2$}
  \put(19,-4){$\scriptstyle d_1$}
  \put( 8,-4){$\scriptstyle u_2$}
  \put(30,-4){$\scriptstyle u_1$}
  \put(20,18){$\scriptstyle \gamma$}
\end{picture}
\begin{picture}(50,20)(0,0)
  \put(0,0){\wedm 4}
  \put(-3,6){$\scriptstyle W$}
  \put(41,6){$\scriptstyle W$}
  \put(19,3){$\scriptstyle d_1$} 
  \put(-3,-4){$\scriptstyle d_2$}
  \put(41,-4){$\scriptstyle d_2$}
  \put(19,-4){$\scriptstyle W$}
  \put( 8,-4){$\scriptstyle u_2$}
  \put(30,-4){$\scriptstyle u_1$}
  \put(19,18){$\scriptstyle \gamma$}
\end{picture}
\\[2\baselineskip]

\begin{picture}(50,20)
  \put(0,0){\wedm 1}
  \put(-3,6){$\scriptstyle W$}
  \put(41,6){$\scriptstyle W$}
  \put(19,3){$\scriptstyle W$} 
  \put(-3,-4){$\scriptstyle d_2$}
  \put(41,-4){$\scriptstyle d_2$}
  \put(19,-4){$\scriptstyle d_1$}
  \put( 8,-4){$\scriptstyle u_2$}
  \put(30,-4){$\scriptstyle u_1$}
  \put( 8,8){$\scriptstyle \gamma$}
\end{picture}
\begin{picture}(50,20)
  \put(0,0){\wedm 3}
  \put(-3,6){$\scriptstyle W$}
  \put(41,6){$\scriptstyle W$}
  \put(19,3){$\scriptstyle W$} 
  \put(-3,-4){$\scriptstyle d_2$}
  \put(41,-4){$\scriptstyle d_2$}
  \put(19,-4){$\scriptstyle d_1$}
  \put( 8,-4){$\scriptstyle u_2$}
  \put(30,-4){$\scriptstyle u_1$}
  \put(30,8){$\scriptstyle \gamma$}
\end{picture}